\documentclass[twocolumn,amsmath,amssymb,longbibliography,aps,prb,reprint,superscriptaddress] {revtex4-2}

\usepackage{graphicx}
\usepackage{amsmath}
\usepackage{bm}
\usepackage{comment}
\usepackage{multirow}
\usepackage{color}
\usepackage{amsfonts}
\usepackage{booktabs}
\usepackage{ulem}

\begin{document}
\title{Topologically non-trivial gap function and topology-induced time-reversal symmetry breaking in a superconductor with singular dynamical interaction}	
\author{Yue Yu}
\author{Andrey V. Chubukov}
\affiliation{School of Physics and Astronomy and William I. Fine Theoretical Physics Institute, University of Minnesota, Minneapolis, Minnesota 55455, USA} 

\begin{abstract}
In many strongly correlated electron systems, non-Fermi liquid behavior and unconventional superconductivity can be viewed as emerging from an effective 4-fermion interaction with a singular frequency dependence. A pairing instability in such a system is qualitatively different from that in a Fermi liquid and generally gives rise to multiple pairing states with topologically distinct gap functions. However, in the systems studied so far, a topologically trivial solution has the lowest energy. Here we show that a repulsive Hubbard-type interaction with a finite cutoff added to a model with a singular dynamical interaction selects, in some parameter range, the theretofore subleading, topologically nontrivial solution. We consider a minimal model that displays this behavior and show that the transformation between the topologically trivial and nontrivial gap functions necessarily occurs via an intermediate phase with topology-induced breaking of time-reversal symmetry. 
\end{abstract}
\maketitle

A non-Fermi-liquid behavior and unconventional superconductivity in strongly correlated electron systems, e.g., fermions at the $\nu =1/2$ Landau level\cite{lee1989,*blok1993,nayak1994,*kim1994,altshuler1994}, clear and dirty quantum-critical metals\cite{Bergeron2012,raghu2015,acf,*acs,*acn,Max_2010,*Max_2010a,*Max_2011,*Max_2015,Lee_2018,Varma2020,Abanov2020,*Abanov2026,Shi_2025,Foster_2023}, quantum dots\cite{Chowdhury2022,esterlis2026,*classen2021,wang2020,*wang2020solvable}, systems near a Mott phase\cite{Georges1996,simard2019,Capone2009,*Capone2023,Held2024,Arovas2022}, are fascinating issues which have attracted substantial interest in recent years in both condensed-matter and high-energy communities. 
Previous studies have shown that in many such systems, the interplay and competition between tendencies towards non-Fermi liquid (NFL) and pairing are adequately described by assuming that the dominant interaction between fermions is dynamical\cite{nayak1994,altshuler1994,raghu2015,esterlis2026,Abanov2020,*Abanov2026,pimenov2022}, with a singular frequency dependence in the infrared limit. This singular interaction gives rise to NFL behavior and mediates pairing out of a NFL in a model-specific non-ordinary $s-$wave channel. Pairing out of a non-Fermi liquid is not associated with the Cooper logarithm, and for this reason, it is qualitatively different from that mediated by a non-singular interaction in a Fermi liquid. An outcome, relevant to this work, is the existence of a tower of topologically distinct solutions of the gap equation\cite{wu2021,*paper5,*paper6,Zhang2022}, in sharp contrast with the conventional BCS/Eliashberg theory in which the gap equation has a single solution.
A solution $\Delta_n$ has $n$ zeros along the Matsubara axis, where $\Delta_n (\omega_m)$ can be set as real by proper choice of the $U(1)$ gauge, or, equivalently, $n$ by $2\pi$ phase slips of the phase of complex $\Delta_n (\omega)$ along the real axis. In more general terms, each zero of $\Delta (z)$ in the upper (causal) half-plane of $z = \omega' + i \omega_m$ can be identified as a dynamical vortex as the phase of a complex $\Delta (z)$ varies by $2\pi$ under an anticlockwise rotation around each such point.
These vortices are global objects and their effects can be detected by time-dependent phase-sensitive Josephson interferometry\cite{wolf2012} and angle-resolved photoemission spectroscopy\cite{vekhter2003,kordyuk2005}. The topological distinction between the gap functions with different $n$ arises because a vortex cannot be created or annihilated within the causal plane; it may enter or leave the causal plane only from its boundaries, which are the real frequency axis and infinity. 
\footnote{In some models, such as the $\gamma$-model of metallic quantum criticality with $\gamma >1$, the behavior is even more complex, as even the gap function $\Delta_0$ with no nodes on the Matsubara axis still possesses dynamical vortices in the upper half-plane of complex frequency.}
For the models studied so far, e.g. $\gamma$-models of metallic quantum criticality~\cite{Abanov2020,*Abanov2026} (and Yukawa-SYK models of quantum dots\cite{esterlis2026,*classen2021,wang2020,*wang2020solvable}), the topologically trivial solution $n=0$ is the ground state, while solutions with $n >0$ are saddle points in the space of system parameters~\cite{Zhang2022}. Given that other solutions are present, it is highly desirable to find the conditions under which a topologically non-trivial solution is the ground state. 

\begin{figure}[h]
\centering
\includegraphics[width=7cm]{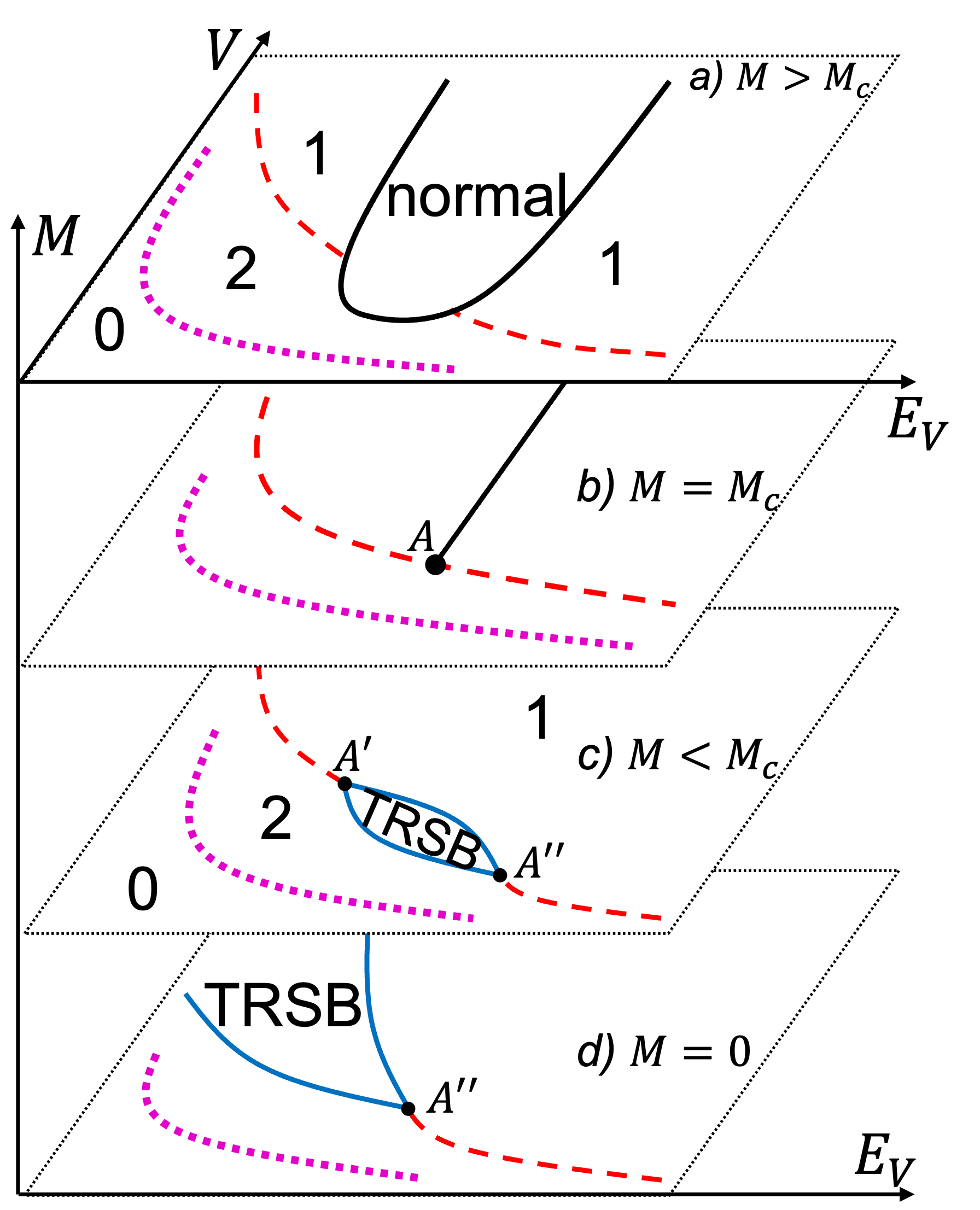}
\caption{Schematic phase diagram for the model with an effective attraction mediated by a dynamical boson with $\chi (\Omega_m) = 1/(|\Omega_m|^\gamma + M^\gamma)$ (the $\gamma$-model) with additional Hubbard-type repulsion $V$ with a hard energy cutoff $E_V$. 
The $\gamma$-model without $V$ has either a single solution for the superconducting gap (as in BCS theory) for $M > M_c$ or two solutions for $M \leq M_c$ (and more solutions for smaller $M$, which we don't consider here). The plots are for different values of $M$ around $M_c$ on the $(E_V,V)$ plane. The numbers $0,1,2$ specify the number of dynamical vortices of the gap function in the causal (upper) half-plane of frequency. For $M > M_c$, when there is only one solution for the gap in the pure $\gamma$-model, there is a parameter range with no superconductivity (labeled as ``normal"). For $M =M_c-0^+$, when the second solution for the gap in the $\gamma$-model emerges, this range shrinks to a line ending at point $A$. For smaller $M$, superconductivity with a finite gap $\Delta (\omega)$ exists everywhere in $(E_V,V)$ plane and breaks TRS in some range of $E_V$ and $V$. The phase diagram at $M=0$ neglects other solutions for the gap function in the $\gamma$-model. Blue solid lines are a second-order phase transition to the time-reversal symmetry-breaking (TRSB) phase. Black solid lines are second-order superconducting transitions. Red dashed lines are topological phase transitions between phases with 2 and 1 vortices in the causal half-plane. Magenta dotted lines are topological phase transitions between phases with 0 and 2 vortices.
}
\label{F0}
\end{figure}

In this paper, we report on our findings on such conditions. We consider for definiteness the $\gamma$-model of quantum criticality, in which the attractive interaction between fermions is provided by a soft boson with mass $M$, whose dynamical susceptibility on the Matsubara axis is $\chi (\Omega_m) = 1/(|\Omega_m|^\gamma + M^\gamma)$, where here and below we measure $\Omega_m$, $M$, and other energy variables in units of fermion-boson coupling, which is the only energy scale in the $\gamma$-model. We add to this model a repulsive interaction $V$ with a hard upper cutoff at energy $E_V$ (controlled in practice by an external gate) and analyze the phase diagram in the $(E_V,V)$ plane. The limits $E_V\gg1$ and $E_V\ll1$ can be viewed as adiabatic and anti-adiabatic \cite{gastiasoro2020}, respectively. 
We use the fact that vortex solutions with $n >0$ disappear one by one as the bosonic mass $M$ increases, beginning with the solutions with the largest $n$, and focus on the range of $M$ where the gap equation without $V$ has at most two solutions, which we label as $\Delta_0$ \& $\Delta_1$. $\Delta_0$ with zero vortex on the Matsubara axis is the only solution for $M > M_c$; and $\Delta_0$ and one-vortex $\Delta_1$ are the two solutions at $M \leq M_c$. We label the gap function of the complete model with $V \neq 0$ as $\Delta$. At small $V$, the gap function $\Delta (z) \approx \Delta_0 (z)$ is topologically trivial. The function $\Delta_0 (z)$ is approximately constant at $|z| <1$ and decays as $1/z^\gamma$ at $|z| \gg 1$. 

Our results are summarized in Fig. \ref{F0}. We argue that a new behavior emerges for larger $V$. Specifically, we show that for $M > M_c$, $\Delta_0$ undergoes a series of topological transitions, and most importantly, there exists a region in the $(E_V,V)$ plane where superconductivity vanishes once $V$ exceeds a certain cutoff (panel a in Fig. \ref{F0}). At $M = M_c$ (panel b), the normal state shrinks to a line, along which an infinitesimally small $\Delta$ emerges with the structure $ \Delta (z) \equiv \Delta_1 (z)$. For $M <M_c$, the gap $\Delta (z)$ is finite everywhere on the phase diagram, but still for certain $E_V = E_{V,c}$, $\Delta (z) \equiv \Delta_1 (z)$ for $V$ larger than some threshold value $O(1)$. For small $V$ and the same $E_V = E_{V,c}$, $\Delta (z) = \Delta_0 (z)$. 
We show that, as the two states are topologically distinct, the transformation between them necessarily occurs via an intermediate phase with broken time-reversal symmetry (TRSB phase) (panels c and d). 
The gap function in this intermediate phase is approximately $\Delta = \cos{\theta} \Delta_0 \pm i \sin{\theta} \Delta_1$, where the sign is chosen spontaneously due to TRSB and $\theta$ varies from $0$ to $\pi/2$ between the lower and upper boundary of the TRSB phase.
We emphasize that in our case, the existence of the TRSB phase is a consequence of the different topological structures of $\Delta_0$ and $\Delta_1$, i.e., TRSB is driven by the topology. In this respect, this case 
is the reverse of the well-studied cases in which topology (e.g., chiral edge modes) is a consequence of TRSB. The TRSB phase is tiny for $M$ slightly below $M_c$ (panel c), but widens up at smaller $M$ (panel d). We show that the TRSB phase emerges via a highly non-trivial phase transition at $M = M_c$. 

To obtain these phase diagrams, we combine numerical analysis with analytical reasoning. We numerically solve the gap equation on the Matsubara axis (Eq.\ref{eq_D}) below, obtain $\Delta (\omega_m)$ in various ranges on the $(E_V,V)$ plane, and use Padé approximants to obtain a complex $\Delta(\omega)$ along the real axis. We verified that the Padé approximation accurately captures the moment at which a vortex crosses from the lower into the upper half-plane. And the solution along the Matsubara axis accurately captures when the vortex enters/leaves this axis, including $\omega_m = \infty$. 

We begin with the case $M > M_c$, where in the absence of $V$, the gap function $\Delta=\Delta_0$ is topologically trivial. The evolution of the gap function with increasing $V$ is the same as in panel (a) in Fig. \ref{F0}. We show the numerical phase diagram in Fig.\ref{F:EM1} in the End Note. 
In the adiabatic regime $E_V \gg 1$, the evolution of $\Delta_0$ with $V$ has previously been studied for electron-phonon interactions\cite{rietschel1983,coleman2015,pimenov2022}, and the behavior in our model is qualitatively similar. Superconductivity persists even at large $V$, but the topology of the gap function undergoes a series of changes: (i) at a critical value $V = V_{c1}$ (magenta dashed line), two dynamical vortices enter the upper half-plane of complex frequency $z$ through the real-axis boundary at $z=\pm\omega$; (ii) at a larger $V$ the two vortices merge on the Matsubara axis and split with one vortex moving up and the other moving down along the Matsubara axis; (iii) at $V = V_{c}$ (red dashed line), the vortex moving up exits the causal plane through its boundary at infinity, leaving a single dynamical vortex on the Matsubara axis. Consequently, the gap function $\Delta(z)$ has zero vortices for $V < V_{c1}$, two vortices for $V_{c1}< V < V_{c}$, and one vortex for $V > V_{c}$. 
In the anti-adiabatic regime $E_V \ll 1$, the number of vortices also evolves with increasing $V$ from zero to two to one, but the gap evolution is different. Namely, at $V = V_{c1}$ (magenta dashed line) the two vortices enter the upper half-plane through the real-axis boundary at $z = \pm\omega$, like in the adiabatic case, however, at a larger $V$, they move towards $z=0$ along the curved trajectory and reach it at $V=V_{c}$ (red dashed line). 
At even larger $V$, one vortex moves up along the Matsubara axis while the other vortex moves into the lower half-plane. In the end, the gap function again has a single vortex in the causal plane, along the Matsubara axis. However, the evolution of this state with increasing $V$ is very different from the one in the adiabatic regime. Namely, in the antiadiabatic regime, the second vortex exits through the origin, while in the adiabatic regime, it exits through infinity. For this reason, one cannot gradually move from one single-vortex state to another by changing $E_V$ at some large $V$. Simple experimentation shows that the only option is the normal state in between. Indeed, we found numerically that the two single-vortex states at large $V$ and large and small $E_V$ and separated by the normal phase at $E_V = O(1)$ (see Fig.\ref{F:EM1}). 

The situation changes at $M =M_c -0^+$. Now the gap equation without $V$ has two solutions: a vortex-free $\Delta_0$ with finite magnitude and an infinitesimally small single-vortex $\Delta_1$. At small $V$, the emergence of the second solution is irrelevant as $\Delta (z) = \Delta_0 (z)$ is non-zero. However, for larger $V$ and $E_V = O(1)$, $\Delta (z)$, viewed as a continuation of $\Delta_0 (z)$, vanishes, as we just showed. We now argue that $\Delta_1 (z)$ does not vanish at a certain $E_V = E_{V}^c =O(1)$, even when $V$ is large. To see this, consider the gap equation on the Matsubara axis in the presence of $V$. It is 
\begin{widetext}
\begin{equation}
\Delta(\omega_m)=\int_{0}^{\infty} \frac{1}{2} 
\frac{d\omega_m'}{\sqrt{\omega_m'^2+|\Delta(\omega_m')|^2}} \left(\frac{\Delta(\omega_m')-\Delta(\omega_m)\frac{\omega_m'}{\omega_m}}{|\omega_m-\omega_m'|^\gamma+M^\gamma} + \frac{\Delta(\omega_m')+\Delta(\omega_m)\frac{\omega_m'}{\omega_m}}{|\omega_m+\omega_m'|^\gamma+M^\gamma} \right)- V \int_0^{E_V}d\omega_m'\frac{\Delta(\omega_m')}{\sqrt{\omega_m'^2+|\Delta(\omega_m')|^2}} 
\label{eq_D}
\end{equation}
\end{widetext}
For infinitesimally small $\Delta (\omega_m) = \Delta_1 (\omega_m)$, ${\sqrt{\omega_m^2+|\Delta(\omega_m)|^2}}$ can be replaced by $|\omega_m|$. We obtained an infinitesimal $\Delta_1 (\omega_m)$ by solving (\ref{eq_D}) in $M = M_c$ and $V=0$. We show the result in Fig.\ref{F:EM1} in the End Note.

The function is relatively small at small $\omega_m$, then changes sign, peaks at $\omega_m = O(1)$, and rapidly decreases at larger $\omega_m$. For such $\Delta_1 (\omega_m)$, the last term in (\ref{eq_D}) can be eliminated by properly choosing $E_V = E_V^c = O(1)$. For this $E_V$, $\Delta_1 (\omega_m)$ is then the solution of (\ref{eq_D}) for any $V$, even large $V$. 
We solved the gap equation (\ref{eq_D}) in $M = M_c -0^+$ for different $V$ and $E_V$ and found that the region of normal state behavior in Fig.\ref{F0}a shrinks to the line $E_V = E_V^c$ (see Fig.\ref{F0}b).
We present the exact phase diagram at $M=M_c-0^+$ in Fig. \ref{F:EM1}. We emphasize that this occurs entirely because at $M =M_c -0^+$ the gap equation at $V=0$ develops a second solution $\Delta_1$.

Analyzing this case even more carefully, we find that the single-vortex solution $\Delta = \Delta_1$ along $E^c_{V}$ terminates at the critical point $A = (E^c_{V},V^*)$. Immediately below this point, $\Delta (z)$ is also infinitesimally small and has two vortices. This $\Delta (z)$ is the continuation of $\Delta_0 (z)$ to $V \leq V^*$. We label it as ${\bar \Delta}_0$. The point $A$ then hosts two infinitesimally small solutions $\Delta_{1}$ and ${\bar \Delta}_0$. We show in the End Note that the frequency dependence of ${\bar \Delta}_0$ is rather special: We find that $\Delta (z)$, taken as a mixture (a linear combination) of
${\bar \Delta}_0$ and $\Delta_{1}$, evolves under anti-clockwise rotation around $A$ (starting from a point below it) from $\Delta (z) ={\bar \Delta}_0$ with two vortices inside the causal plane to mixed $\Delta (z)$ with one 
vortex at a finite Matsubara frequency and one at infinity (red line on the right from A), then to $\Delta (z) = \Delta_1 (z)$ with one vortex 
again along the Matsubara axis and finally to mixed $\Delta (z)$ with two vortices at the origin (red line on the left from A). Note that while the shape of the gap functions changes dramatically around $A$, the free energy 
is continuous around $A$ because the magnitude of $\Delta (z)$ is infinitesimally small.

\begin{figure}[h]
\centering
\includegraphics[width=0.9 \linewidth]{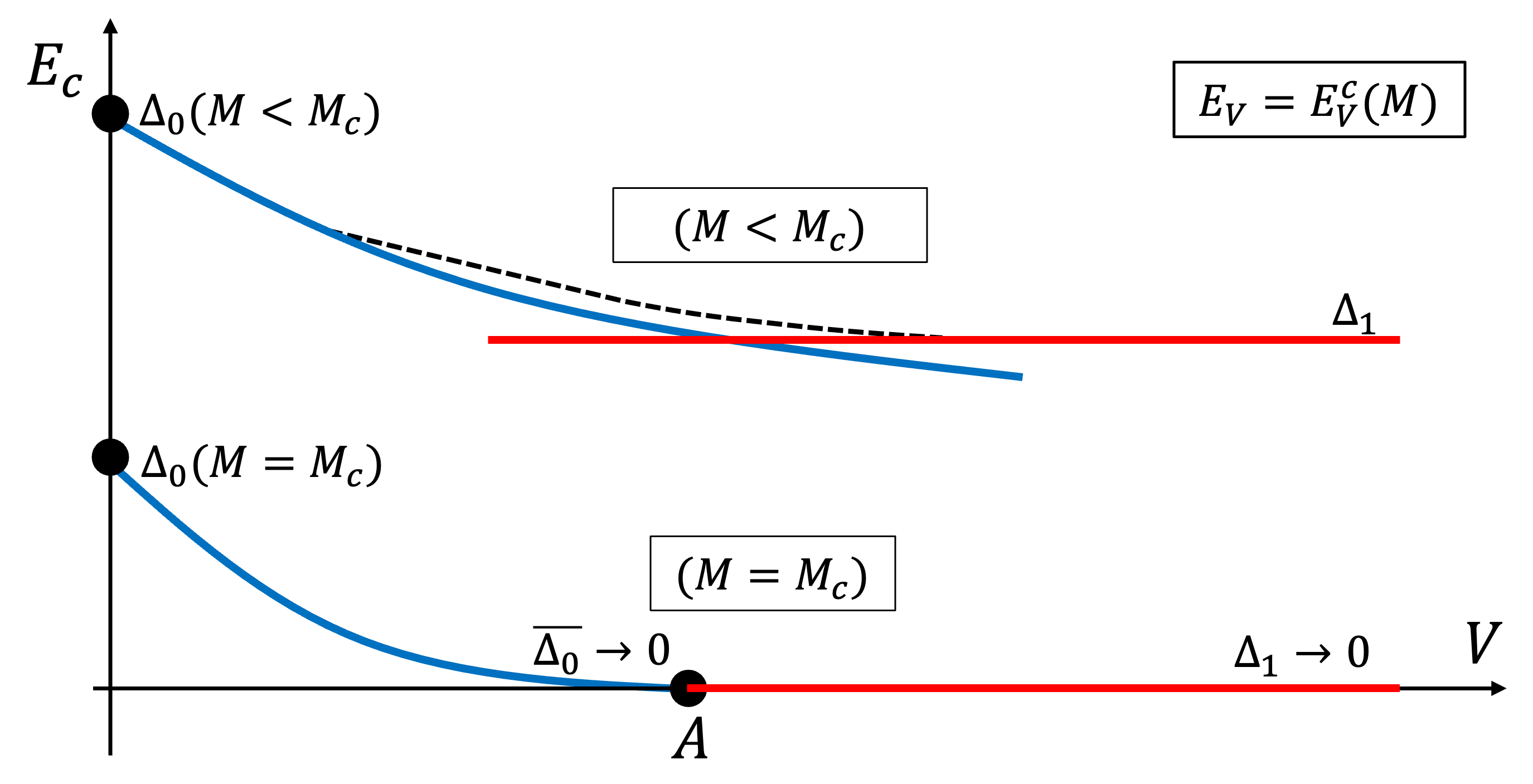}
\caption{Condensation energy $E_c$ as a function of V at $E_V=E_V^c$ for $M=M_c$ and $M<M_c$. The red line is $E_c$ for the large $V$ solution $\Delta = \Delta_1$ and the blue line is 
$E_c$ for the small $V$ solution $\Delta = {\bar \Delta}_0$. 
The black dashed line is $E_c$ in the TRSB phase. For $M =M_c$, $E_c$ vanishes at $V = V^*$ (the $A$ point) and remains zero at larger $V$ as for $M = M_c$, $\Delta_1$ is infinitesimally small. }
\label{F3}
\end{figure}

Next, we move to $M < M_c$. We still have $\Delta \equiv \Delta_1$ at $E_V = E^c_V$ at $V > V^*$, but now $\Delta_1$ has a finite magnitude. 
\footnote{The values of $E^c_V$ and $V^*$ depend on $M$, i.e., $E^c_V = E^{c}_V (M)$ and $V^* = V^{*}(M)$. We skip the index $M$ to simplify the presentation.}.
At the same $E_V = E^c_V$ and $V < V^*$, $\Delta = {\bar \Delta}_0$ also has finite magnitude. 
However, these two gap functions cannot continuously evolve into each other at $V^*$ as they have a different number of vortices ($1$ and $2$) and are therefore topologically distinct.
We analyze analytically and numerically how to resolve this and find that the resolution is an intermediate TRSB phase where the gap function is complex:
$\Delta \approx \cos\theta\Delta_1+i\sin\theta{\bar \Delta}_0$ and $\theta$ varies from $0$ to $\pi/2$ between $V > V^*$ and $V < V^*$. Extending the analysis to other values of $E_V$, we find that the TRSB phase occupies a finite parameter range, the area of which increases as $M$ decreases.
In particular, along the $2\to1$ transition line (red line in Fig.\ref{F0}), the critical point $A$ splits into $A'$ and $A''$, which become the end points of the TRSB phase. We show the numerical phase diagram in Fig.\ref{F:EM2}. 

\begin{figure}[h]
\centering
\includegraphics[width=0.95\linewidth]{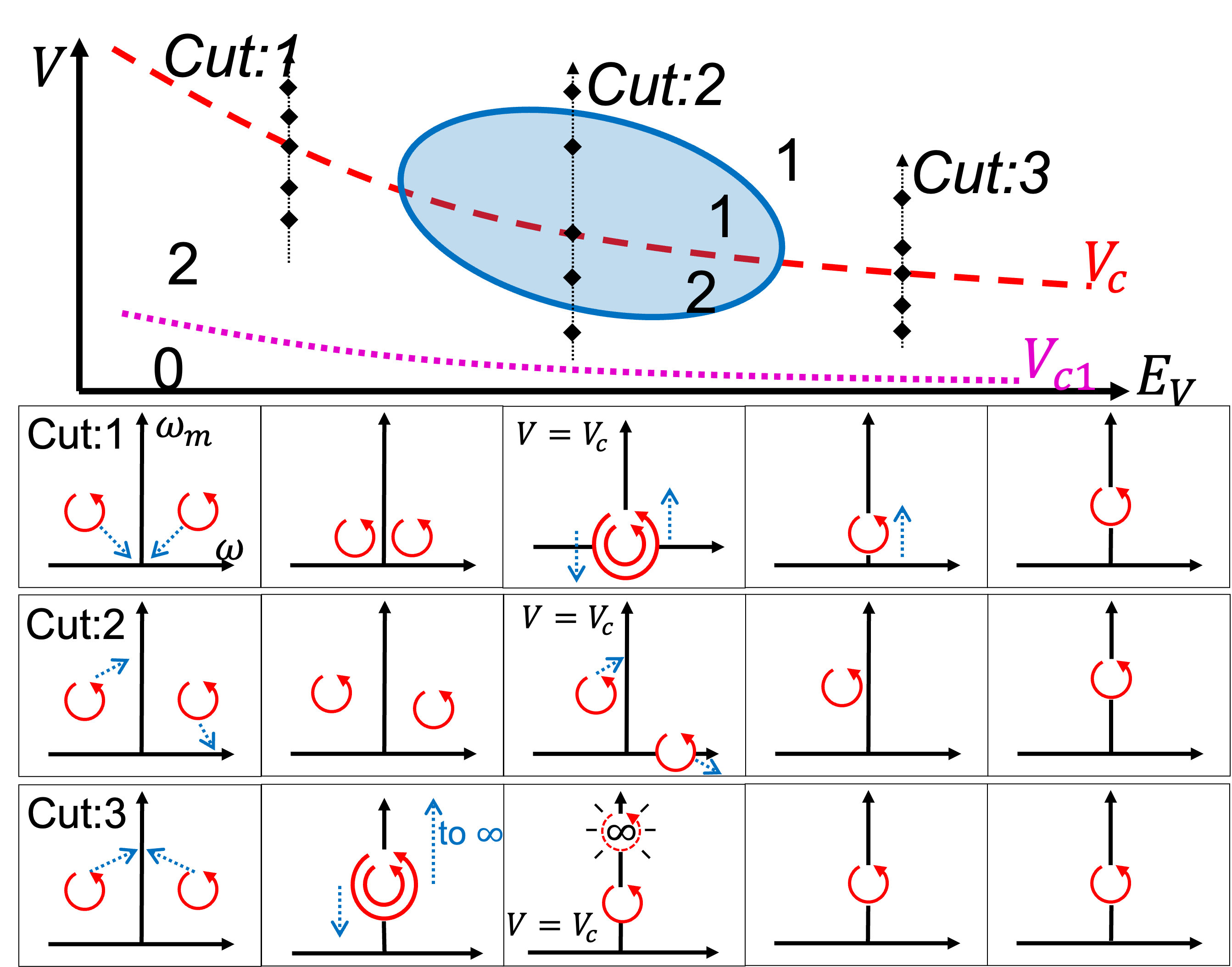}
\caption{
The evolution of the vortex structure of the gap function along three different cuts at small, intermediate, and large $E_V$. 
Red circles indicate vortices; the direction of the arrow indicates that $2\pi$ phase winding is for anticlockwise rotation. Along the line labeled as $V_c$, the number of vortices in the causal (upper) half-plane of frequency changes from 2 at $V < V_c$ to 1 at $V > V_c$. At small $E_V$ (cut 1), one of two vortices leaves the causal half-plane at the origin, at large $E_V$ (cut 3), it leaves at infinity. To smoothly connect the two limits, one of two vortices has to leave the causal plane at some finite $\omega$ on the real axis. This is only possible if TRS is broken at intermediate $E_V$, because otherwise, when one vortex leaves a causal half-plane at $\omega$, another must leave it simultaneously at $-\omega$. If TRS is broken (cut 2, TRSB phase is shaded), there is no symmetry between $\omega$ and $-\omega$, and a single vortex can leave the causal half-plane at a finite $\omega$. 
}
\label{F1}
\end{figure}

The TRSB phase is topologically protected by the topological distinction between the behavior along the $2\to1$ transiton lines to the left and to the right of the TRSB (Fig.\ref{F1}). Upon
crossing this line at small $E_V$, a vortex exits the causal plane at the origin. 
Upon crossing it at large $E_V$, a vortex exits at infinity. To continuously connect these two topologically distinct limits, a single vortex must exit the causal plane at a finite real frequency. This is only possible when the TRS symmetry between $z$ and $-z^*$, i.e., between
a vortex on the left side of the causal half-plane and a vortex on the right side, is broken such that the two vortices evolve differently under a variation of $V$. In this situation, one of the two vortices can exit the causal half-plane through the real axis, changing the number of vortices in the causal half-plane from 2 to 1. 
\footnote{TRS maps $\Delta(\omega + i\omega_m)$ to $\Delta^*(-\omega + i\omega_m)$. As it flips both $\omega$ and Im($\Delta$), a vortex on the left side of the causal plane becomes a vortex on the right side. In the TRS
phase, a gap function with an odd number of vortices in the causal half-plane must have an odd number of vortices on the Matsubara axis.} As illustrated in Fig.\ref{F1}, there is generically a transition line inside the TRSB phase between $\Delta(z)$ with 2 and 1 vortices. 

To locate the boundaries of the TRSB phase, we depart from the TRS-symmetric states with 
one or two vortices on the Matsubara axis and numerically compute the matrix of the second derivatives of the condensation energy with respect to all possible variations of the initially real $\Delta (\omega_m)$. We find that the leading instabilities are towards the appearance of an imaginary component of $\Delta (\omega_m)$, which implies TRSB. 
We solve the gap equation for a complex $\Delta (\omega_m)$ and compute the condensation energy along the $E_V = E^c_V$ line. We find (see Fig.\ref{F3} and Supplementary Information\cite{supp}) 
that the TRSB state indeed has the largest negative condensation energy between $A'$ and $A''$.

In Fig.\ref{F:EM2}, we show the phase diagram at $M=0$ assuming that other solutions for the gap function in the pure $\gamma$-model, $\Delta_n (\omega_m)$ with $n >1$,
are irrelevant in the range of $V$ and $E_V$ that we consider. The TRSB phase is wider at $M=0$, and its lower end $A'$ is at $E_V =0$ and $V = \infty$. 

To put our results in a broader perspective, we note that in our model, the topologically protected TRSB phase
is introduced to smoothly connect two distinct $2\to1$ transition lines.
In general, the TRSB phase can appear near the phase transitions between $2n$ and $2n+1$ vortices.
In the $\gamma$-model for $\gamma>1$, $\Delta_0$ already contains $2n$ vortices in the upper half-plane, where $n$ varies with $\gamma$ (no vortices on the Matsubara axis), and $\Delta_1$ contains an additional vortex on the positive Matsubara axis, and the set of transitions with increasing $V$ are from $2n$ to $2n+2$ to $2n+1$ vortices. 

To conclude, in this work, we address two issues: (i) can one detect 
topologically distinct solutions $\Delta_n$ for pairing mediated by a near-critical boson, and (ii) can a topological distinction between these solutions induce a phase with broken TRS? 
We use the $\gamma$-model of quantum-critical pairing in a metal and propose a general strategy for realizing a phase diagram in which (i) the subleading, topologically non-trivial gap function $\Delta_1$ appears at the exact ground state solution on a certain line in a parameter space and (ii) topological distinction between $\Delta_0$ and $\Delta_1$ induces a topologically protected time-reversal symmetry-breaking phase in some parameter range. 
The strategy is to add Hubbard repulsion $V$ with a cutoff $E_V$ and vary both $V$ and $E_V$. 
Our approach can be straightforwardly generalized to other models of pairing mediated by a singular dynamical interaction. 
The three parameters $E_V, V$, and the bosonic mass $M$ are all experimentally accessible. In particular, the cutoff scale $E_V$ is set by screening and can be modified by varying the Fermi energy.
Placing the system between metallic gates allows one to control both $E_V$ and $V$. 
A boson mass $M$ can be tuned by pressure or doping, which controls the distance to a critical point.

We conclude with a few remarks on the TRSB state. 
In real space, TRSB induces a current loop order. Since we assume 
a single-momentum-component gap function, our TRSB order does not break the crystal mirror symmetry, and the net orbital angular momentum $L_z\sim xp_y-yp_x$ vanishes.
However, the local orbital angular momentum can remain finite, inducing a staggered current loop order. For lattice models, the current loop order can also preserve discrete translational symmetry if the lattice is bipartite. Such a mirror-even current loop order is sometimes referred to as an orbital altermagnetic state\cite{yu2025,pan2025}. It has symmetry $k_xk_yL_z$ in orthorhombic systems, as opposed to the (spin) d-wave altermagnet $k_xk_yS_i$\cite{vsmejkal2022}. Experimentally, $\mu$SR\cite{sonier2000} can detect the local magnetic fields generated by this staggered current loop order. For global probes, one may apply strain $\epsilon_{xy}$ to explicitly break the mirror symmetry to induce a net orbital angular momentum $L_z$, which does not depend on spin-orbit coupling and can be quite large. This angular momentum can then be measured, for example, by the Kerr effect\cite{kapitulnik2009}, with the Kerr angle varying linearly with the applied strain.

We acknowledge with thanks the useful discussions with Ar. Abanov, E. Berg,
S. Sachdev, J. Schmalian, Y. Wang, Y. Wu and S-S Zhang. 
The work by AVC was supported by the NSF DMR-2325357.
The authors acknowledge support from the Simons Foundation grant SFI-MPS-NFS-00006741-07 for the Simons Collaboration on New Frontiers in Superconductivity.

\bibliography{citation}

\clearpage

\onecolumngrid
\section{End Matter}
In this section, we present the actual phase diagrams, which we obtain in our numerical analysis of the $\gamma$-model 
with bosonic mass $M$ and additional Hubbard repulsion $V$ with the energy cutoff at $E_V$. We present the phase diagrams for different values of $M$, both above and below the critical $M_c$, below which the second solution of the gap equation in the pure $\gamma$- model emerges. To obtain the phase diagrams, we need to know the solution of the complete non-linear gap equation, 
Eq.\ref{eq_D}, on the Matsubara axis. To solve this equation, we discretize the frequency domain using a logarithmic grid: $N$ points are taken from $\omega_{min}$ to $\omega_{max}$, and another $N$ points from $-\omega_{max}$ to $-\omega_{min}$.

\begin{figure}[h]
\centering
\includegraphics[width=0.45\linewidth]{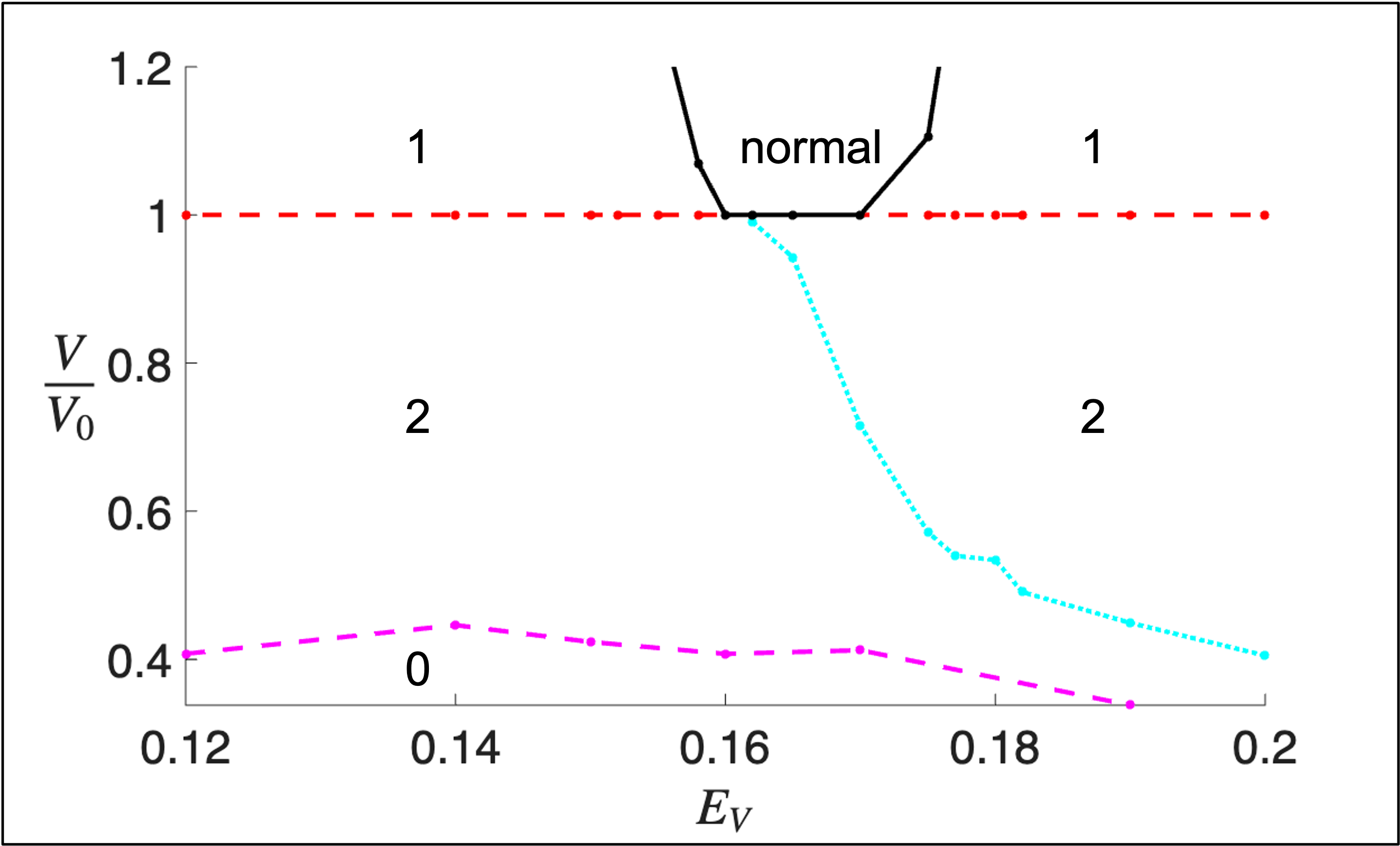}
\includegraphics[width=0.45\linewidth]{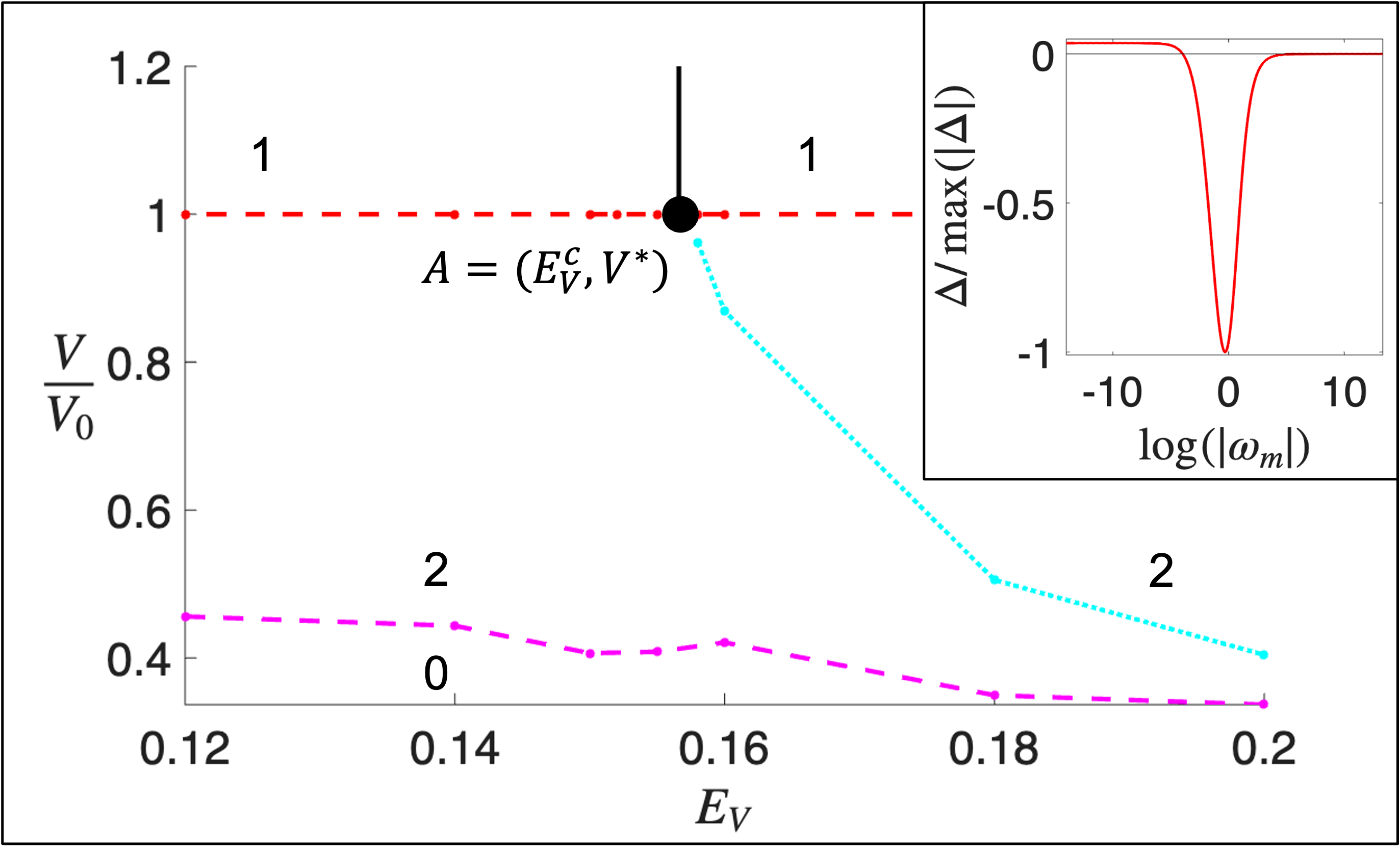}
\caption{The phase diagrams for $\gamma=1.5$, which we obtained numerically. For this $\gamma$, $M_c\approx0.08$. The topologically trivial solution $\Delta_0 (z)$ has two vortices in the causal plane, away from the Matsubara axis.
These two vortices do not leave the causal plane upon increasing $V$, nor do they move to the Matsubara axis, and do not play a role in our analysis.
Numbers in the phase diagram specify the number of {\it additional} vortices in the gap function, relative to that in $\Delta_0$. The magenta dashed line marks the topological phase transition between phases with 0 and 2 vortices. The red dashed line marks a topological phase transition between phases with 2 and 1 vortices. The cyan dotted line specifies where the two vortices in the phase labeled as $2$ merge onto the Matsubara axis to form a double vortex (this is not a phase transition line). The black solid line is the boundary of the normal phase. 
We normalize $V$ by $V_0$ chosen such that the dashed red line is at $V/V_0 =1$.
(Left panel) Phase diagram for $M=0.09>M_c$. The $2\to1$ transitions at small and large $E_V$ are separated by a normal phase.
(Right panel) Phase diagram for $M=M_c-0^+$. The normal phase shrinks to a line, 
ending at a critical point $A=(E_V^c,V^*)$. 
On this line, the (infinitesimally small) gap function is the second solution of the gap equation in the pure $\gamma$-model, $\Delta=\Delta_1 (z)$. We show this solution in the inset, in arbitrary units, 
for a representative $V > V^*$.}
\label{F:EM1}
\end{figure}

\begin{figure}[h]
\centering
\includegraphics[width=0.45\linewidth]{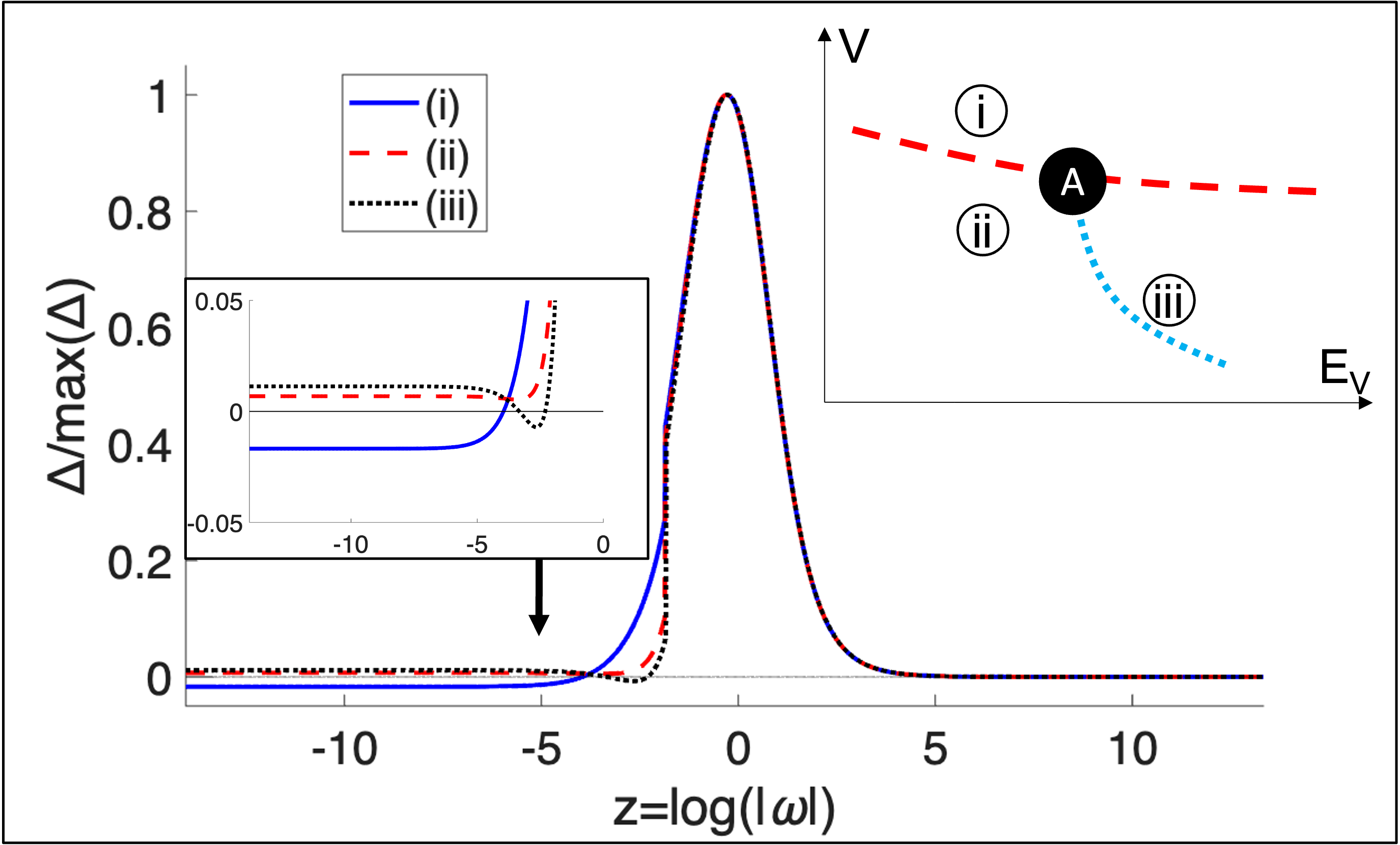}
\includegraphics[width=0.45\linewidth]{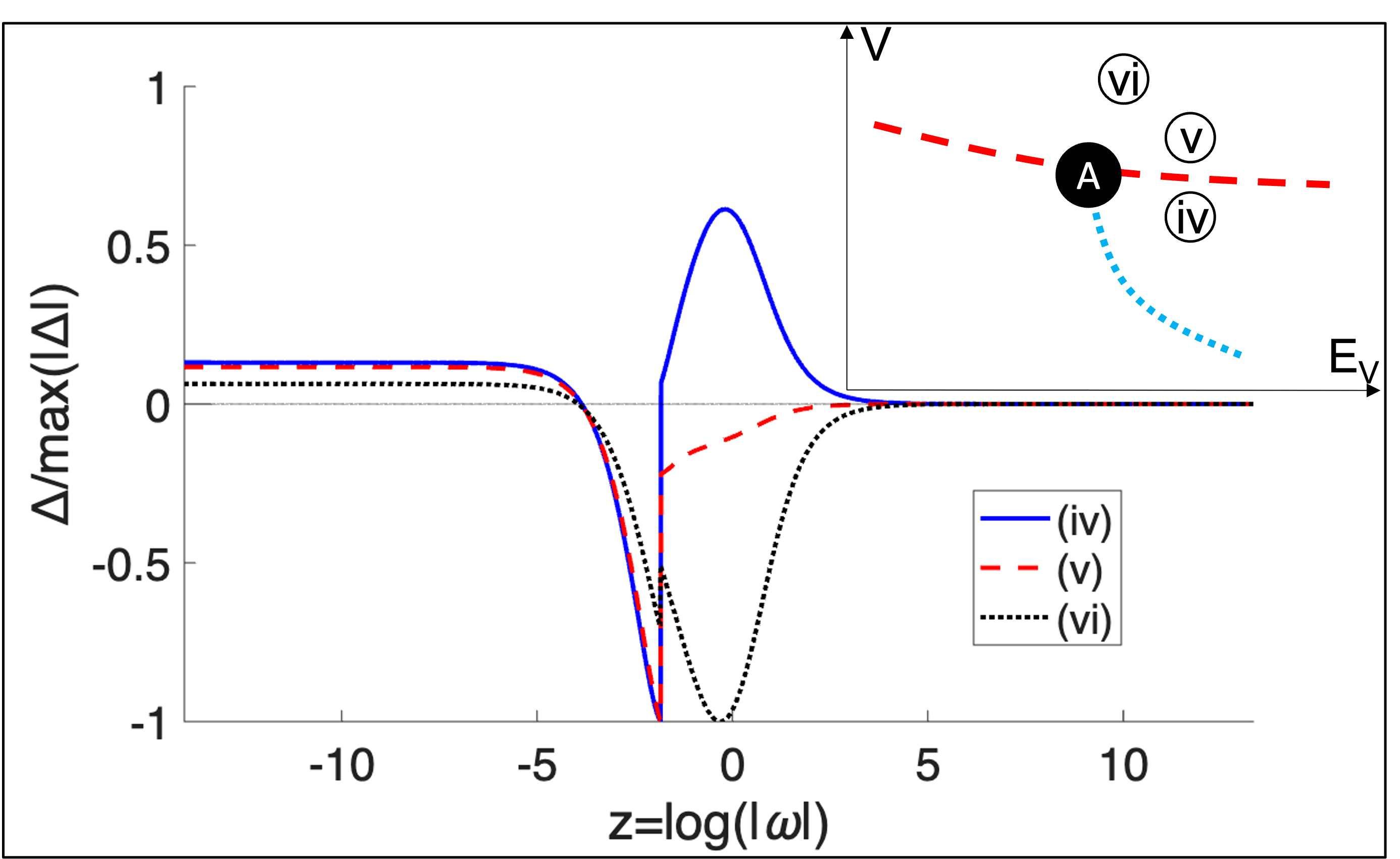}
\caption{The gap function around the point $A=(E_V^c,V^*)$ $M\approx M_c$, obtained from the numerical solution of the gap equation. The points around $A$, for which the solutions are presented, are shown in the inset figures at the top-right corners. Left and right panels -- the evolution of $\Delta (\omega_m)$ on the Matsubara axis
in a counterclockwise rotation, starting at point I and ending at point VI. The gap 
$\Delta (\omega_m)$ at different points is shown in different colors, explained in the legend. 
The gap at points I-III is shown on the left panel, and the gap at points IV-VI on the right panel.
At point I $\Delta (\omega_m)$ (blue solid line) has one vortex at a small Matsubara frequency (a zoom-in figure at small $\omega_m$ (large negative $z = log{|\omega_m|}$) is included). Going counter-clockwise towards the $1\to 2$ transition, this vortex moves to the origin, merging with another vortex from $\omega_m<0$ to form a double vortex at $\omega_m =0$ ($\Delta (\omega_m)$ at this point scales as $\omega^2_m$ at the smallest $\omega_m$). 
At point ii (red dashed line), $\Delta(\omega_m)$ is sign-preserving, indicating that the two vortices moved away from the Matsubara axis, into a causal plane. Moving towards point iii, these two vortices merge on the Matsubara axis at a cyan dotted line and then split along the Matsubara axis. 
At point iii (black dotted line), $\Delta (\omega_m)$ has two vortices (two nodal points).
At point iv (blue solid gap), the two vortices move further away along the Matsubara axis. 
Upon crossing the red dashed line, one vortex moves to infinity, and at point V (red dashed line), only one vortex remains on the Matsubara axis.
At point vi (black dashed line), $\Delta (\omega_m)$ becomes almost the same as $\Delta_1 (\omega_m)$ shown in the inset in the right panel of 
Fig. \ref{F:EM1}. Note that $\Delta (\omega_m)$ is very small and flat for both positive and negative $z = \log{|\omega_m|}$. This allows nodal points to move fast along the Matsubara axis.}
\label{F:EM3}
\end{figure}

\begin{figure}[h]
\centering
\includegraphics[width=0.45\linewidth]{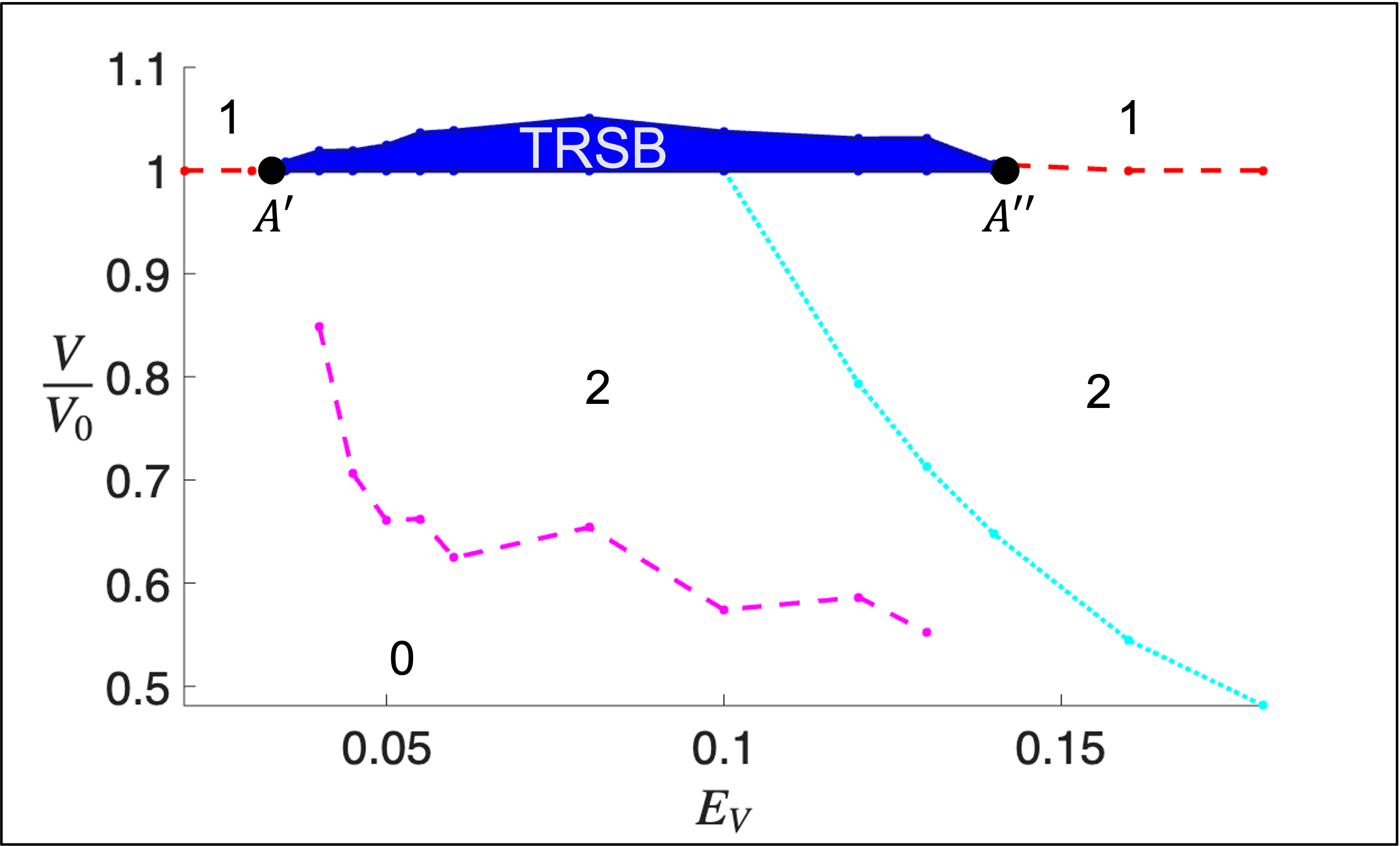}
\includegraphics[width=0.45\linewidth]{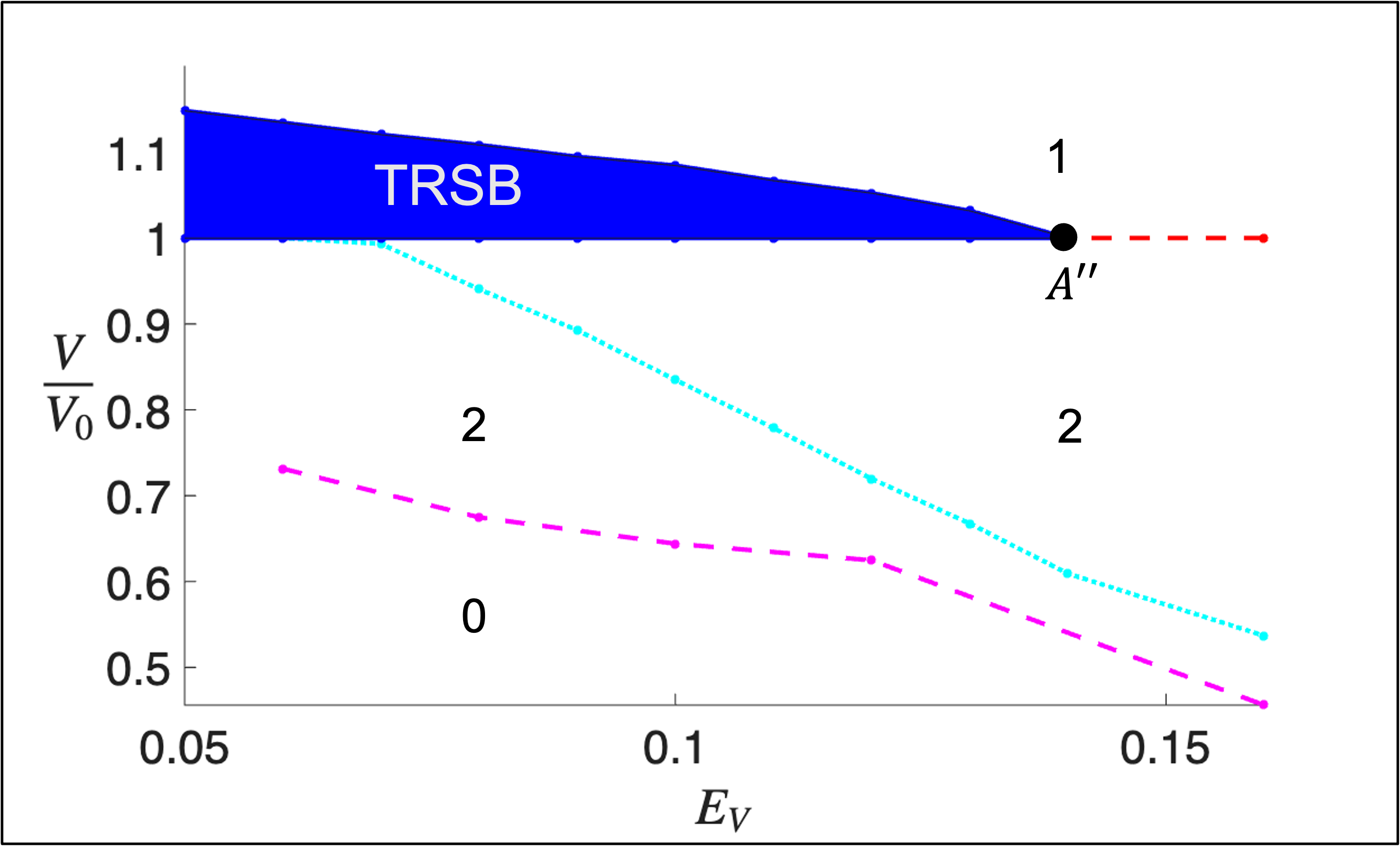}
\caption{The numerical results for $M=0.02 < M_c$ (Left panel) and $M =0$ (Right panel). The notations are the same as in Fig.\ref{F:EM1}. The phase with a spontaneous TRSB is in the blue region. 
As the boson mass $M$ decreases, the area of the TRSB region expands from a point $A$ at $M = M_c -0$ to a range between $A'$ and $A''$ points along $V = V_0$ in the left panel. The boundaries of the TRSB region are second-order phase transition lines, upon crossing which the gap function develops an imaginary component along the Matsubara axis. At $M=0$, the TRSB phase extends to $E_V=0$. We don't show the line along which the $2\to1$ vortex transition occurs inside the TRSB phase.}
\label{F:EM2}
\end{figure}

We choose $N=2000$, $\omega_{min}=e^{-20/\gamma}$, and $\omega_{max}=e^{20/\gamma}$. For convenience of calculations, we set $\gamma =1.5$. For this $\gamma$, 
$M_c \approx 0.08$. For this $\gamma$, even the leading solution of the pure $\gamma$-model, $\Delta_0 (z)$, the one which we label topologically trivial in the main text, actually has two vortices at $z = \pm\omega' + i \omega_m$ in the causal plane, away from the Matsubara axis. As $V$ increases, the system first undergoes a topological phase transition 
from a state with 2 vortices to a state with 4 vortices and then, at larger $V$, from a state with 
4 vortices to a state with 3 vortices.
As we said in the main text, the presence of two vortices in $\Delta_0 (z)$, which are not on the Matsubara axis and do not move critically upon increasing $V$, does not affect our analysis (see Sec. D in the Supplementary Material (SM)).
Our phase diagram depends on the number of {\it additional} vortices, generated by the Hubbard interaction and, at $M < M_c$, on the existence of the second solution for the gap function in the pure $\gamma$-model, $\Delta_1 (z)$, with a node on the Matsubara positive half-axis. For consistency with the analysis in the main text, we label the solution $\Delta_0 (z)$ as topologically trivial and label only the number of additional vortices. In other words, the label ``0" denotes the phase with the structure of the gap function isomorphic to $\Delta_0 (z)$. In Fig.\ref{F:EM1} we show the phase diagrams for $M > M_c$ ( $M=0.09$) and $M \approx M_c$. In Fig.\ref{F:EM3} we show gap functions around $A=(E_V^c,V^*)$ for $M\approx M_c$. The hard cutoff in the Hubbard interaction leads to a discontinuity at $\omega_m=E_V$, which does not affect discussions of vortex motion to the origin or infinity. In Fig.\ref{F:EM2} we show the phase diagrams for $M < M_c$ ($M =0.02$) and $M = 0$.

Some technical details are presented in the SI, which we split into Secs. A-E.

In Sec. A, we analyze the behavior of $\Delta_1(M)$ as $M\to M_c$. In Sec. B, we focus on the condensation energy $E_c$. We analytically show that for $M<M_c$, $E_c$ is minimized at the $E_V=E_V^c$ and large $V$, to the condensation energy of $\Delta_1$. This confirms the lack of a normal phase for $M<M_c$, and the appearance of a normal phase on $E_V=E_V^c$ and $V>V^*$ at $M=M_c$.
In Sec C, we show the details of the decomposition of the gap function in the TRSB phase into 
$\Delta=\cos\theta\Delta_1+i\sin\theta\overline{\Delta_0}$. 
In Sec. D, we discuss the details of the Padé approximation analysis for the $0\to 2$ transition. In 
Sec. E, we show that a direct second-order phase transition between a normal phase and a TRSB superconducting phase is allowed within Landau theory in a 3D phase diagram. 
In Sec. F, we discuss the $\mathbb{Z}_2$ topological invariant for the topologically protected TRSB phase.

\renewcommand{\thesection}{\Alph{section}}
\clearpage
\appendix
\onecolumngrid
\setcounter{figure}{0}
\renewcommand{\thefigure}{SI-\arabic{figure}}
\section{$\gamma$-model}
The  Eliashberg equations for the pairing vertex and the self-energy ($\widetilde{\Sigma}={\Sigma}+\omega_m$)
at $T=0$ are:
\begin{equation}
	\begin{split}
		&\Phi(\omega_m)=\int d\omega_m'\frac{\Phi(\omega_m')}{\sqrt{\widetilde{\Sigma}^2(\omega_m')+|\Phi(\omega_m')|^2}}\times\left(\frac{g^\gamma}{2}\frac{1}{|\omega_m-\omega_m'|^\gamma+M^\gamma}-\frac{V}{2}|_{\omega_m,\omega_m'<E_V}\right)\\
		&\widetilde{\Sigma}(\omega_m)=\omega_m+\int d\omega_m'\frac{\widetilde{\Sigma}(\omega_m')}{\sqrt{\widetilde{\Sigma}^2(\omega_m')+|\Phi(\omega_m')|^2}}\times\frac{g^\gamma}{2}\frac{1}{|\omega_m-\omega_m'|^\gamma+M^\gamma}.
	\end{split}
\end{equation}
The model involves four energy scales: the attraction strength $g$, the repulsion strength $V$, the repulsive cutoff $E_V$, and the boson mass $M$. In this work, we work in units where $g=1$. The superconducting gap function $\Delta=\omega\Phi/\widetilde{\Sigma}$ is shown in Eq.(1). The free energy is
\begin{equation}
	\begin{split}
		F&=-\int d\omega_m\frac{\omega_m^2}{\sqrt{\omega_m^2+|\Delta(\omega_m)|^2}}-\frac{1}{4}\int d\omega_m d\omega_m'\frac{\omega_m\omega_m'+\Delta^*(\omega_m)\Delta(\omega_m')}{\sqrt{\omega_m^2+|\Delta(\omega_m)|^2}\sqrt{\omega_m'^2+|\Delta(\omega_m')|^2}}\times\frac{1}{|\omega_m-\omega_m'|^\gamma+M^\gamma}\\&+\frac{V}{4}\left|\int_{-E_V}^{E_V}d\omega_m\frac{\Delta(\omega_m)}{\sqrt{\omega_m^2+|\Delta(\omega_m)|^2}}\right|^2
	\end{split}
\end{equation}

\begin{figure}[h]
	\centering
	\includegraphics[width=8cm]{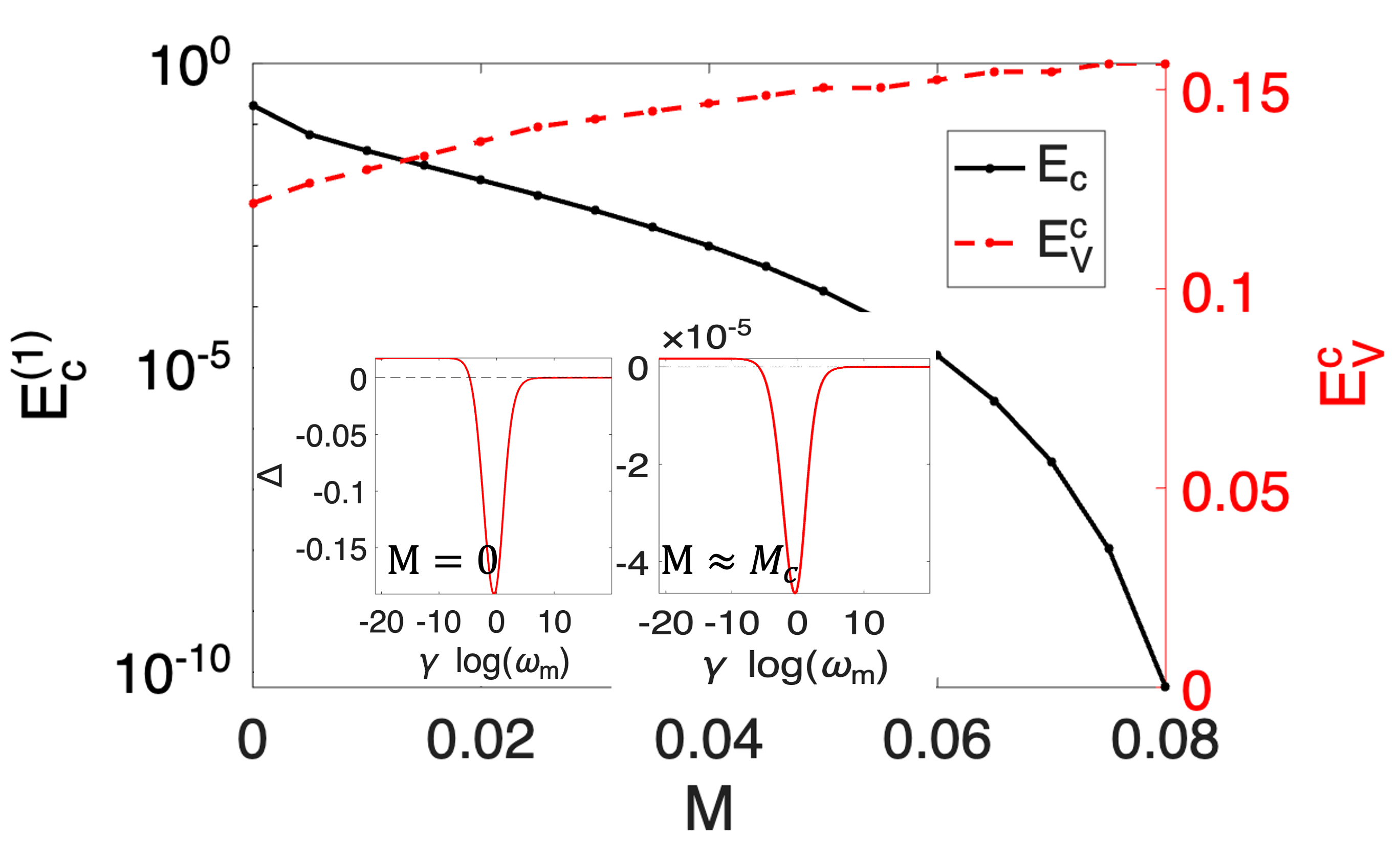}
	\caption{Condensation energy $E_c^{(1)}$ of the subleading solution $\Delta_1$ in the $V=0$ problem as a function of boson mass. The mass weakly affects the shape of the gap function and the critical $E_V^c$ for which repulsive energy vanishes. Inset is the gap function at $M=0$ and $M=0.08$. $\gamma=1.5$ is used.}
	\label{F:SI1}
\end{figure}

At $V=0$, the nonlinear gap equation can have multiple $\omega_m$-even solutions. Since the interaction is purely attractive, the leading solution $\Delta_0$ is sign-preserving. A subleading solution $\Delta_1$ exists for $M \leq M_c$ and features a single vortex (i.e., a sign change) along the positive Matsubara axis, as shown in Fig.\ref{F:SI1}. Higher-order solutions containing more vortices can appear at even smaller $M$; however, they are not relevant for this work. These solutions are topologically distinct, as they carry different numbers of vortices in the causal plane.

\section{Minimum of condensation energy}
\begin{figure}[h]
	\centering
	\includegraphics[height=5cm]{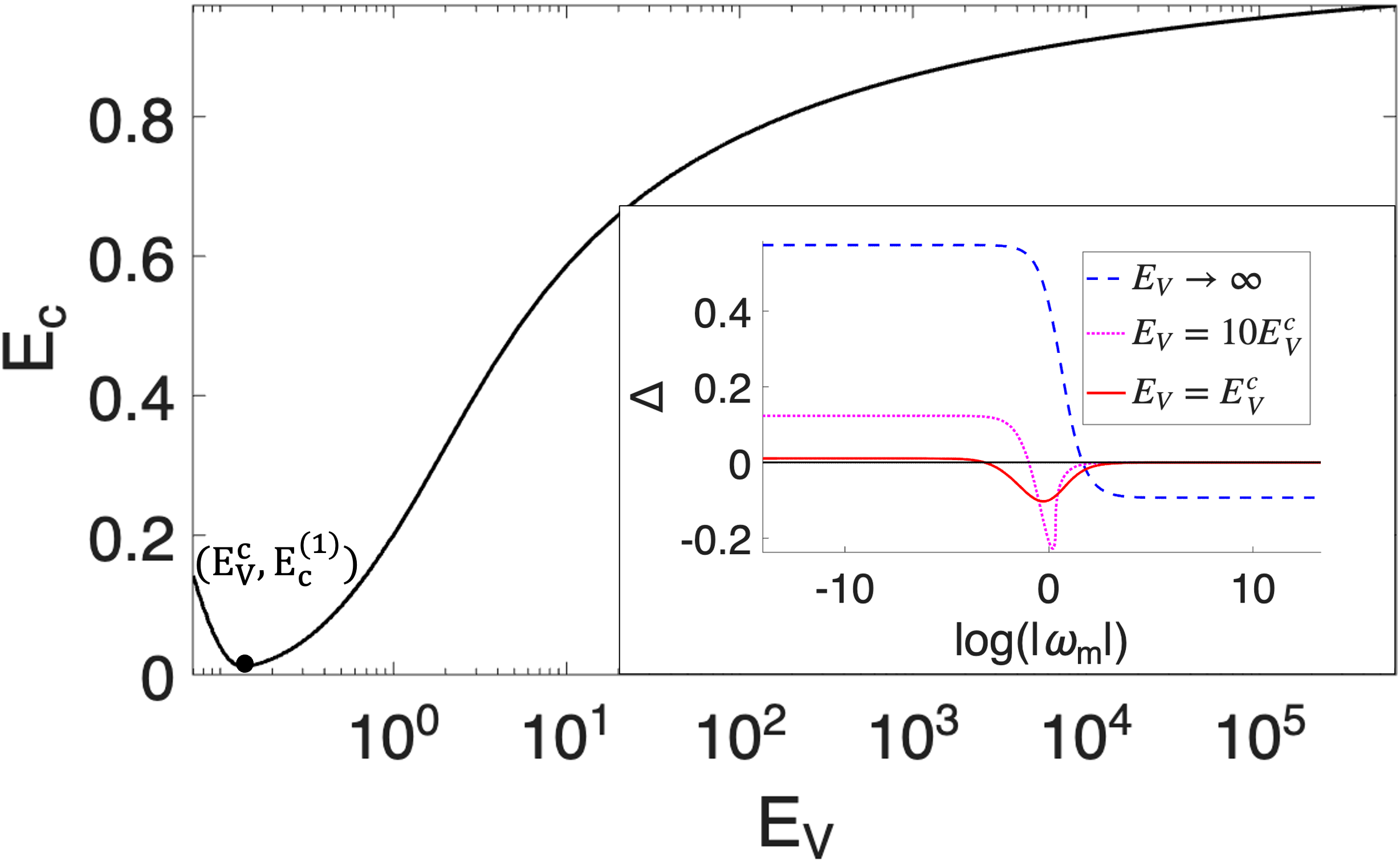}
	\includegraphics[height=5cm]{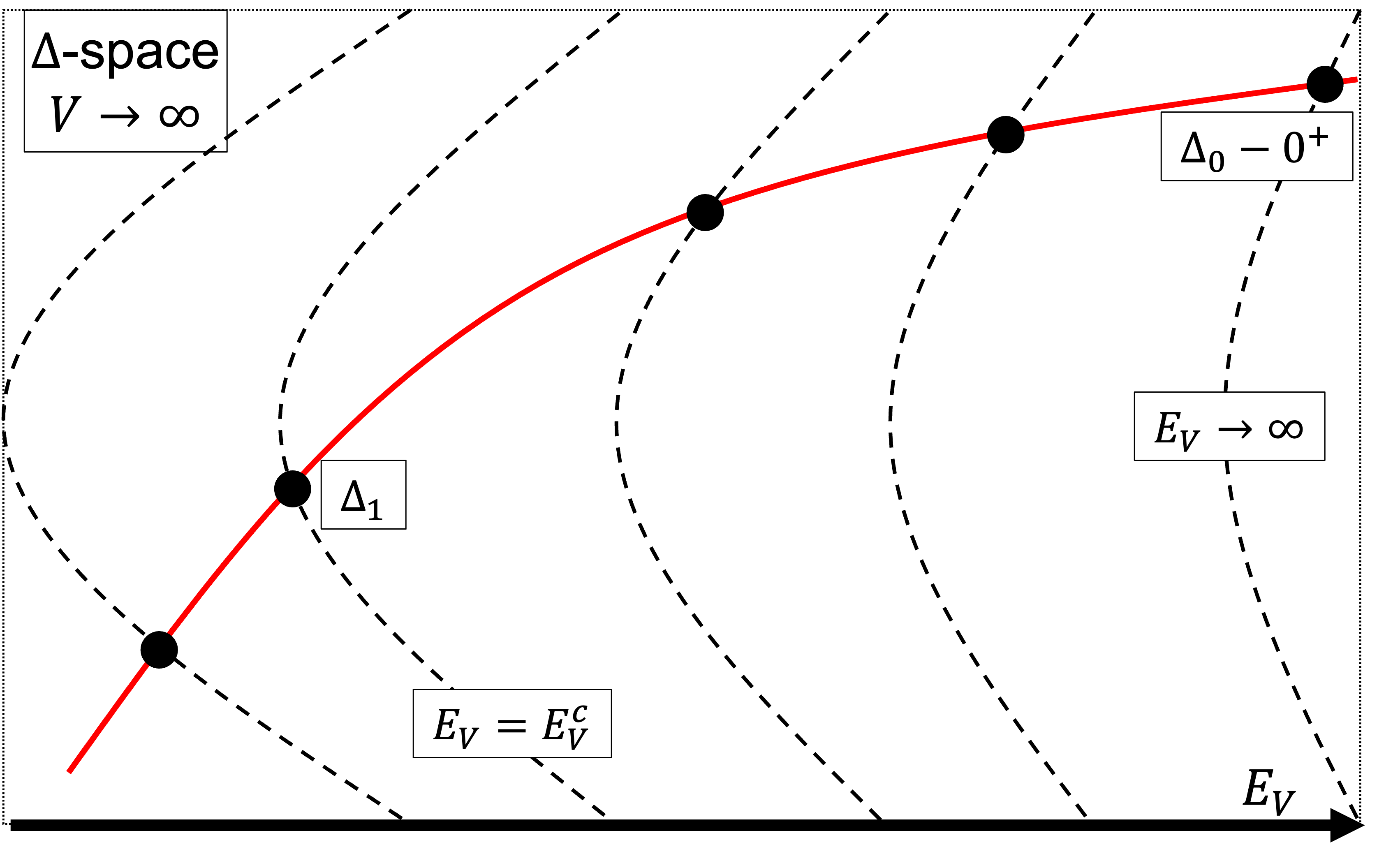}
	\caption{(Top) Condensation energy $E_c$ as a function of $E_V$ at large $V$. Ground state gap functions at three different $E_V$ are included. The vortex moves to a lower frequency as $E_V$ is reduced. The minimal $E_c$ corresponds to the subleading solution $\Delta_1$ of the $V=0$ problem. (Bottom) Solutions at $V\rightarrow\infty$ and various $E_V$ form the 1D red line in the $\Delta$-space. Each solution lives in a subspace (black-dashed line) that minimizes the repulsive energy $F_V$ with the given $E_V$. $M=0.02$ is used. }
	\label{F:1}
\end{figure}

Here we prove that, for $M<M_c$, the condensation energy is minimized at $E_V^c$ and large $V>V^*$. We focus on the condensation energy at the strong repulsion limit $V \rightarrow\infty$. This phase features a single vortex on the positive Matsubara axis. Due to the large $V$, ground state solutions all minimize the repulsive energy (second line) $F_V$ at their $E_V$, by having a vortex on the Matsubara axis. To minimize $F_V$, the location of the sign-change is bounded by the limit of integration $E_V$.  
For $E_V\rightarrow\infty$, the solution $\Delta\approx\Delta_0^-$ has a vortex at infinity. As $E_V$ decreases, this vortex moves toward lower frequencies, as shown in the top panel of Fig.\ref{F:1}.

We compare the condensation energies $E_c$ for these $E_V$-dependent ground state solutions. Since they have all minimized the repulsive energy $F_V$, it is sufficient to compare the attractive contribution (denoted as $F_0$). This is related the $V=0$ problem, where $\Delta_{0,1}$ both satisfy the gap equation $\delta F_0/\delta \Delta=0$. Here, $\Delta_0$ is the global minimum of $F_0$, while the subleading $\Delta_1$ is unstable towards $\Delta_0$. In $\Delta$-space (right panel of Fig.\ref{F:1}), ground states at different $E_V$ form a 1D path (red line), passing through $\Delta_1$ at $E_V=E_V^c$. Each solution lives in a subspace (black dashed line) that minimizes the repulsion for the given $E_V$. Since $\Delta_1$ satisfies the gap equation $\delta F_0/\delta \Delta=0$, its condensation energy $E_c^{(1)}$ is a local extremum on the 1D red path. $E_c^{(1)}$ is the minimum, as the red path goes towards $\Delta_0^-$. The numerical results are in the left panel.

We now have $\Delta_1$ as the ground-state solution in the strong-repulsion regime. Its condensation energy $E_c^{(1)}$ is the lowest in the $V\rightarrow\infty$ limit. Consequently, $E_c^{(1)}$ provides a lower bound on the condensation energy at weaker repulsion and throughout the entire $(E_V, V)$ phase diagram. This proves the lack of a normal phase for $M<M_c$. Furthermore, when $M$ increases to $M=M_c$, the normal phase will appear on the line $E_V=E_V^c(M=M_c)$ and $V>V^*$.

\section{gap function in the TRSB phase}
Here, we show the complex gap function $\Delta(\omega_m)$ in the TRSB phase. $M=0$ and $E_V=0.08$ are used in this calculation. The condensation energy of the TRSB phase (black line) is lower than the two real TRS-preserved solutions (green \& red lines).
\begin{figure}[h]
	\centering
	\includegraphics[width=0.5\linewidth]{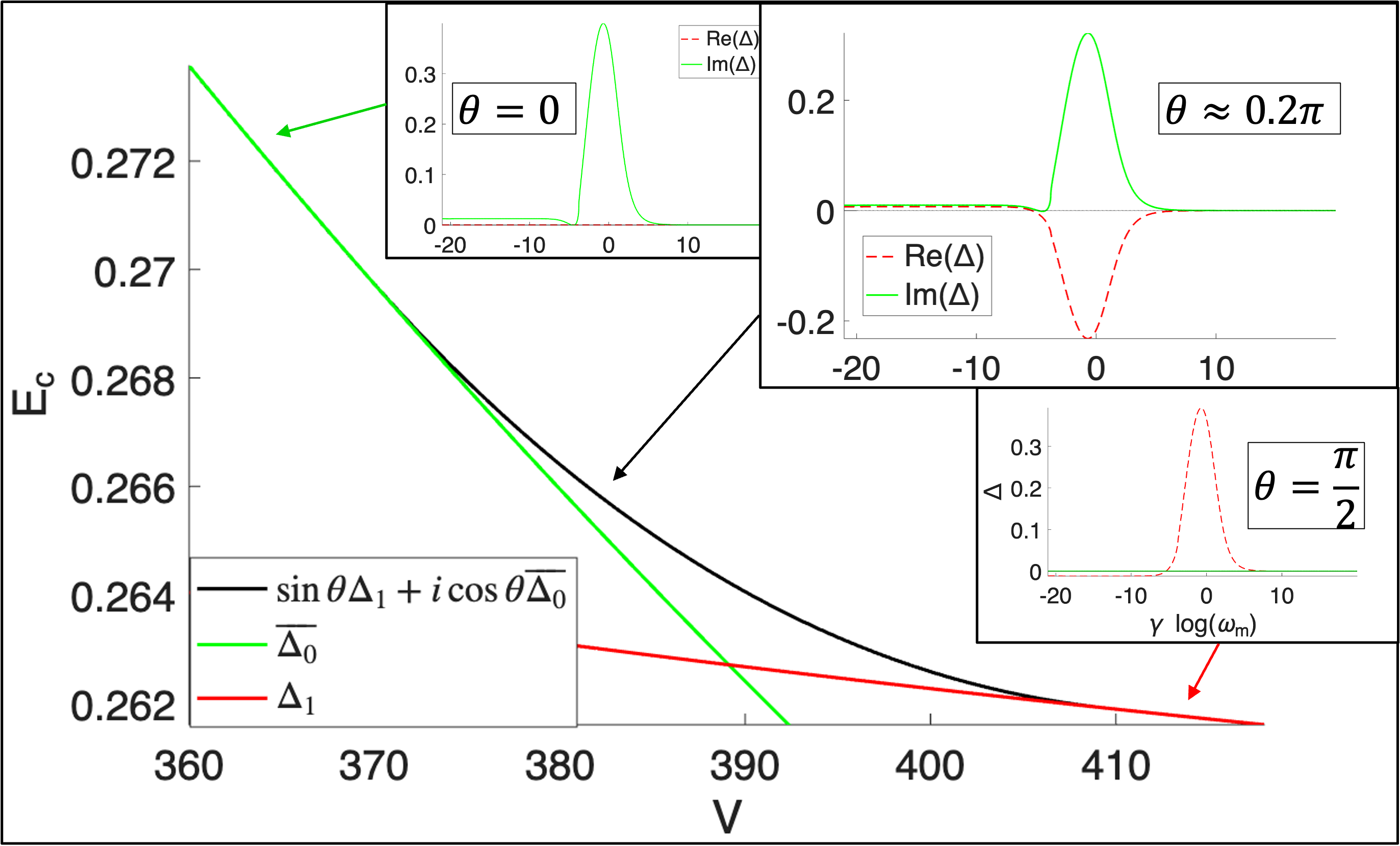}
	\caption{In the TRSB phase, as $V$ increases, the gap function can be approximated by $\Delta=\cos\theta\Delta_1+i\sin\theta\overline{\Delta_0}$, where $\Delta_1$ and $\overline{\Delta_0}$ are the two real solutions at the upper and lower boundaries of the TRSB phase.}
\end{figure}

\section{analytical continuation}
We can determine the number of vortices in the causal plane by analytically continuing the gap function to the real axis and examining its phase winding along the real axis. This gives the $V_{c1}$ for $0\to 2$ transition.

\begin{figure}[h]
	\centering
	\includegraphics[width=8cm]{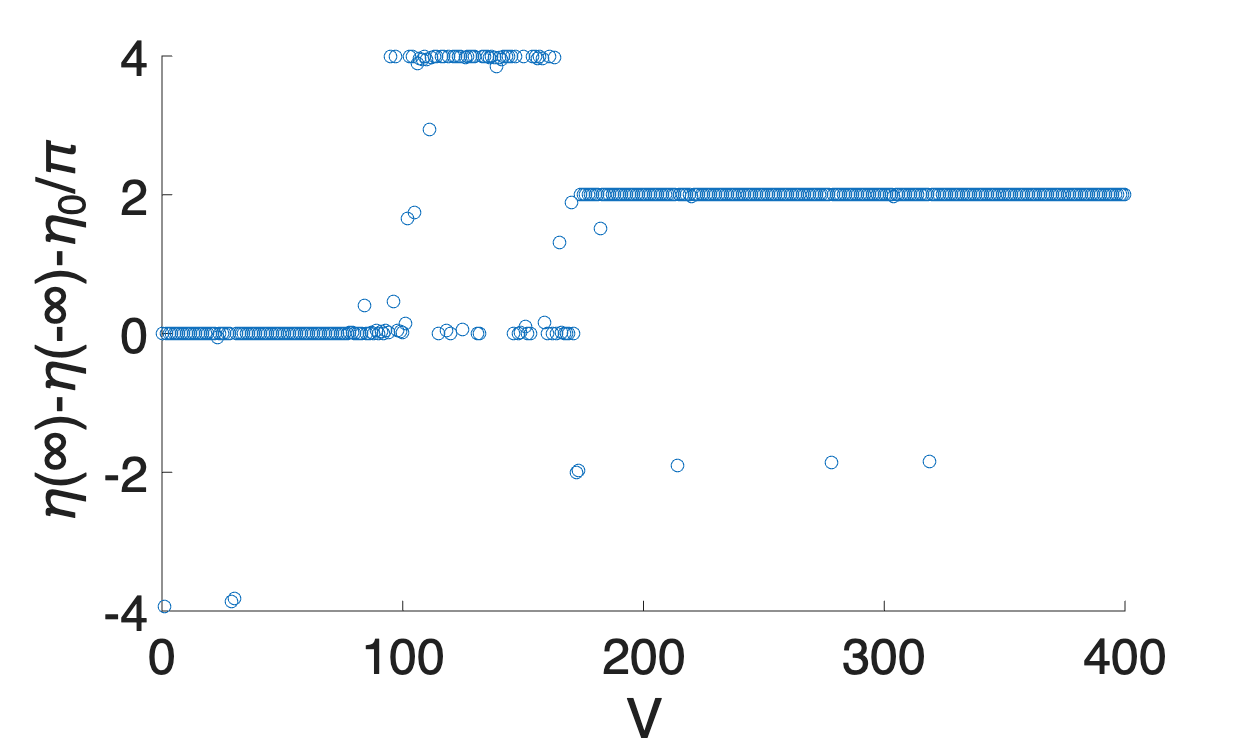}
	\caption{Phase winding along real frequency axis for $E_V=0.14$ for $M=0$ and $\gamma=1.5$. $\eta_0$ is the phase winding at $V=0$.}
	\label{F:A1}
\end{figure}

As an example, we show the analytical continuation for $M=0$ and $E_V=0.14$. Time-reversal symmetry is preserved for all $V$, and the gap function is real on the Matsubara axis. We use a recursive Padé approximation, $\Delta(z)=a_0/(1+a_1(z-z_1)/[1+a_2(z-z_2)/\{1+...\}])$, to fit the gap function on the Matsubara axis. We then evaluate the analytically continued gap function above the real axis, $\Delta(z=\omega+i0^+)$, and extract the phase winding $\eta(\infty)-\eta(-\infty)$. The result is shown in Fig.\ref{F:A1}. We find three distinct phases, with phase windings of $0$, $4\pi$, and $2\pi$, corresponding to $0$, $2$, and $1$ vortices in the upper frequency half plane, measured with respect to the trivial solution $\Delta_0$. The first $0\to2$ transition produces the magenta dashed line in the phase diagram, while the second $2\to1$ transition matches the red dashed line obtained from the analysis on the Matsubara axis.

\section{Landau theory}
Here, we present the Landau theory for the direct transition from the normal phase into the TRSB SC phase with a 1D SC irreducible representation. The free energy density is:
\begin{equation}
	\begin{split}
		f&=\alpha_1|\Psi_1|^2+\alpha_2|\Psi_2|^2+\alpha_3(\Psi_1^*\Psi_2+c.c.)+u_1|\Psi_1|^4+u_2|\Psi_2|^4+u_3|\Psi_1|^2(\Psi_1^*\Psi_2+c.c.)+u_4[(\Psi_1^*)^2\Psi_2^2+c.c.] 
	\end{split}
\end{equation}
Not all quartic terms are included. Here, we introduce redundancy by using two order parameters $\Psi_1$ and $\Psi_2$ to describe the same 1D irreducible representation. This redundancy leads to the $\alpha_3$ bilinear coupling and the $u_3$ coupling.

In a 1D phase diagram, a direct transition from the normal phase into the TRSB phase is forbidden. Without loss of generality, we consider a SC phase transition with $\alpha_1<\alpha_2$. In the SC phase, when $\Psi_1$ is developed, $\Psi_2\propto\alpha_3\Psi_1/\alpha_2$ will be induced by the $\alpha_3$ term. This requires $\Psi_1$ and $\Psi_2$ to have the same phase.

The direct transition is also forbidden in a 2D phase diagram. With an additional parameter to tune, we can now take $\alpha_3=0$. Consider a SC transition $0=\alpha_1<\alpha_2$. When $\Psi_1$ is developed, a small $\Psi_2\propto u_3|\Psi_1|^2\Psi_1/\alpha_2$ will be induced by the $u_3$ term. This also requires $\Psi_1$ and $\Psi_2$ to have the same phase.

The direct transition is allowed in a 3D phase diagram. The SC transition is at $\alpha_1=\alpha_2=\alpha_3=0$. Consider the transition along the line $\alpha_1=\alpha_2=\alpha$ and $\alpha_3=0$. In the SC phase, we can define $|\Psi_1|^2+|\Psi_2|^2\equiv\epsilon^2$. Now the quadratic terms in the free energy density only depend on $\alpha$ and the magnitude $\epsilon$, but not the detailed form of $(\Psi_1,\Psi_2)$. The form of $(\Psi_1,\Psi_2)$ is then determined by the quartic terms. These quartic terms all scale as $\epsilon^4$, so their relative magnitude matters. For example, the competing solutions $(\Psi_1,\Psi_2)=\epsilon(\cos\phi,\sin\phi)$ and $\epsilon(\cos\phi,i\sin\phi)$ can be distinguished by the $u_4$ term. If $u_4$ is large and negative, $(\Psi_1,\Psi_2)=\epsilon(\cos\phi,\sin\phi)$  wins, and time-reversal symmetry is preserved at the SC transition. If $u_4$ is large and positive, the TRSB solution $(\Psi_1,\Psi_2)=\epsilon(\cos\phi,i\sin\phi)$ will win. The exact value of $\phi$ can be determined by minimizing the free energy under all allowed quartic terms.  Since three equations ($\alpha_1=\alpha_2=\alpha_3=0$) are solved for the direct transition, the solutions are point-wise in the 3D phase diagram.
In the main text, the direct transition is at $(E_V^c,V^*)$ of $M=M_c$. Here, $M=M_c$ says that the subleading solution is barely ordered, meaning $\alpha_2=0$. $V=V^*$ says the two solutions are degenerate, meaning $\alpha_1=\alpha_2$. On $E_V=E_V^c$, $\Delta_1$ is fixed and does not mix with $\Delta_0$. This corresponds to $\alpha_3=0$, where the two order parameters are decoupled at the quadratic level in the free energy density. 

While the Landau theory can describe a TRSB transition, it does not tell why the TRSB phase is necessary, because it cannot capture any topological phase transitions involving dynamical vortices.

\section{Topological invariant}
It is instructive to contrast our topological protection with that in a topological insulator. There,  topologically protected edge modes appear between two topologically distinct phases of matter in {\it real} space. The topological invariant is $\mathbb{Z}_2$: materials with invariants 0 and 2 are equivalent, meaning that two edge modes can hybridize and gap out. In our case, the topologically protected TRSB phase arises between two topologically distinct transitions, at small and large $E_V$, in {\it phase} space. The topological invariant is also $\mathbb{Z}_2$: TRSB phase is protected near a $2n+2\to2n+1$ transition, but not near any $2n\to 2n+2$ transitions. Imagine a $0\to 2$ transition (in some models), where in one limit two vortices entered the causal plane from the origin, while in another limit they entered from infinity. These two limits could then be continuously interpolated without passing through an intermediate TRSB phase. In the intermediate regime, the two vortices would instead enter symmetrically from the real axis at positions $\pm \omega$.
\end{document}